\documentstyle[osa,aps,prl,epsfig]{revtex}
\begin{document}
\twocolumn[\hsize\textwidth\columnwidth\hsize\csname
@twocolumnfalse\endcsname
\title{
Observation of abrupt first-order metal-insulator transition in
GaAs-based two-terminal device}
\author{Hyun-Tak Kim$^{\ast}$, Doo-Hyeb Youn, Byung-Gyu Chae, and Kwang-Yong Kang}
\address{Telecom. Basic Research Lab., ETRI, Daejeon 305-350, Republic of Korea}
\author{Yong-Sik Lim$^{\ast\ast}$}
\address{Department of Applied Physics, Konkuk University, Chungju, Chungbuk 380-701, Republic of Korea}
\date{August 24, 2004}
\maketitle{}
\begin{abstract}
An abrupt first-order metal-insulator transition (MIT) as a jump
of the density of states is observed for Be doped GaAs, which is
known as a semiconductor, by inducing very low holes of
approximately $n_p\approx$5$\times$10$^{14}$~cm$^{-3}$ into the
valence band by the electric field; this is anomalous. In a higher
hole doping concentration of
$n_p\approx$6$\times$10$^{16}$~cm$^{-3}$, the abrupt MIT is not
observed at room temperature, but measured at low temperature. A
large discontinuous decrease of photoluminescence intensity at
1.43 eV energy gap and a negative differential resistance are also
observed as further evidence of the MIT. The abrupt MIT does not
undergo a structural phase transition and is accompanied with
inhomogeneity. The upper limit of the temperature allowing the MIT
is deduced to be approximately 440K from experimental data. The
abrupt MIT rather than the continuous MIT is intrinsic and can
explain the "breakdown" phenomenon (unsolved problem) incurred by
a high electric field in semiconductor devices.

PACS numbers: 71.27. +a, 71.30.+h
\\
\end{abstract}
]
\section{INTRODUCTION}
In two-dimensional (2D) systems such as
Si-metal-oxide-semiconductor field-effect-transistors (Si-MOSFETs)
and GaAs/AlGaAs heterostructures, theoretical calculations for
electrical conductivity predicted that these systems were not
expected to be conducting, even though interaction between
carriers is weak (or absent) or very strong
\cite{Abrahams2,Altshuler,Tanatar}. The theoretical predictions
appeared to be proved by experiments where thin metallic films and
Si-MOSFETs displayed faithful logarithmic increase in resistivity
\cite{Dolan} and Si-MOSFET with low electron densities also showed
exponential increase in resistivity \cite{Uren}. However, a
metal-insulator transition (MIT) was observed below 1 K in
GaAs/AlGaAs heterostructures \cite{Gold,Yoon} and in
low-disordered Si-MOSFETs and GaAs/AlGaAs heterostructures
\cite{Pudalov,Shashkin}. As an explanation of the MITs at
extremely low temperature, it was suggested that non-Fermi-liquid
states can induce perfect conductors only when interactions
between electrons are sufficiently strong \cite{Chakravarty}.
Other possible explanations were also given in a review paper
\cite{Abrahams}.

Recent studies on the MIT for GaAs samples such as dilute 2D GaAs
systems have been focused to analyze characteristics of
resistivity and spin-susceptibility at low temperatures
\cite{Zhu,Lilly,Leturcq}. At high temperature such as room
temperature, GaAs samples with lower doping concentrations
exhibited the behavior of thermally activated conduction, while
those samples with higher doping concentrations showed the
metallic conduction feature \cite{Noh}. The latter revealed a
continuous MIT as hole doping concentration was increased.

The MITs observed in 2D systems up to now are continuous for low
voltage and temperature excitations, which doesn't account
necessarily for the possibility of breakdown of the energy gap
because there is the linear metallic regime of conductivity caused
by impurity levels at low applied voltages \cite{Kim1}. However,
an abrupt MIT rather than a continuous MIT seems to be more
intrinsic \cite{Kim1}. An abrupt MIT can exist when a continuous
MIT is true. The abrupt MIT has not been observed in
semiconductors such as low-disordered Si-MOSFETs and GaAs/AlGaAs
heterostructures. The long unsolved MIT problem in GaAs can be
consistently resolved when not only a continuous MIT but also a
abrupt MIT are simultaneously observed.

As a similar research, an abrupt MIT in the representative
strongly correlated material, VO$_2$, was measured and accompanied
with the structural phase transition (SPT) at 67$^{\circ}$C
\cite{Morin}. It was also reported that the abrupt MIT induced by
electric field does not undergo SPT and is accompanied by
inhomogeneity \cite{Kim1,Youn,Chae}. The mechanism was also
revealed by the hole-driven MIT theory describing the abrupt MIT
and inhomogeneity \cite{Kim3,Kim2}. Furthermore, the principal
idea of the abrupt MIT used in VO$_2$ gives a hint to investigate
whether an abrupt MIT in GaAs can occur.

In this paper, we observe an abrupt first-order MIT by inducing
internal hole charges of about
$n_p\approx$5$\times$10$^{14}$~cm$^{-3}$ in hole levels into the
valence band with electric field in GaAs-based two-terminal
devices, on the basis of the hole-driven MIT theory (or extended
Brinkman-Rice picture) and the Mott criterion \cite{Yslim}, as
shown in Fig. 1. The key idea is doping of a very low hole
concentration in GaAs, which indicates an excitation of holes in
the hole levels produced by dopants into the valence band by
electric field. This research focuses on observation of the abrupt
MIT irrespective of theoretical analysis because a continuous MIT
has already been observed in GaAs \cite{Noh}. Furthermore, the
breakdown phenomenon as an unsolved problem in semiconductor
physics is discussed in light of the abrupt MIT. Note that any
band theory cannot self-consistently account for this abrupt MIT
in GaAs, which may be rather explained by strong correlation.
Therefore, GaAs of a band insulator with 8 electrons in $sp^3$
orbital is inevitably assumed as a Mott insulator such as VO$_2$
\cite{Kim1} and is applied to a hole-driven MIT theory to reveal
the abrupt MIT, even if the assumption can be a contradiction in
theory; this is left as an open problem.

\begin{figure}
\vspace{-1.0cm}
\centerline{\epsfysize=12cm\epsfxsize=8cm\epsfbox{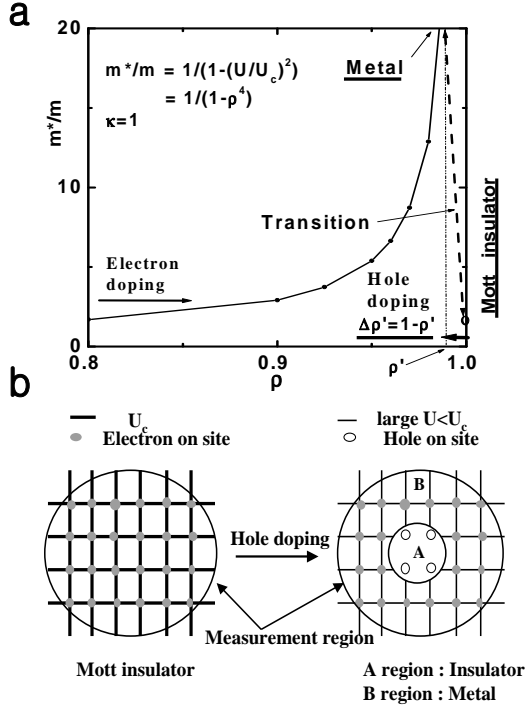}}
\vspace{-0.5cm} \caption{$\bf a$, Curve is the effective mass of
quasiparticle,
$\frac{m^{\ast}}{m}=\frac{1}{1-(U/U_c)^2}=\frac{1}{1-{\kappa^2\rho^4}}$
with $\kappa$=1 and band filling $\rho\ne$1. When $\rho\ne1$ for
an inhomogeneous system, the effective mass is the effect of
measurement and an average of the true effective mass in the
Brinkman-Rice picture (see reference 20,21,23). Electric
conductivity, $\sigma\propto(\frac{m^{\ast}}{m})^2$ (see reference
24). Mott transition occurs just below $\kappa\rho$=1 by hole
doping of $\triangle\rho'$ (low concentration). When Mott
criterion, $n_c^{1/3}a_0\approx$0.25 (see reference 24) is used,
$n_c\equiv\triangle\rho'<$0.001$\%$
$\approx$3.3$\times$10$^{16}~$cm$^{-3}$ (see reference 22) to the
number of carriers in the half-filled band is calculated. $\bf b$,
When low concentration holes, $\triangle\rho'$, are doped in a
Mott insulator created by the critical on-site Coulomb energy of
$U_c$ (the thick line of left side), a breakdown of Coulomb energy
occurs from $U_c$ to a large constant U ($<U_c$)(the thin line of
right side); this is the Mott transition. The system becomes
inhomogeneous due to doped holes (the right side of Fig. 1b). The
metal of region B follows the Brinkman-Rice picture. This is the
hole-driven MIT theory (extended Brinkman-Rice picture).}
\end{figure}

\section{Experiments}
Two-terminal devices were patterned by a lift-off process with Be
doped p-type GaAs films deposited on GaAs substrates by molecular
beam epitaxy, as shown in Fig. 2. Doped hole concentrations of
$n_p\approx$5$\times$10$^{14}$~cm$^{-3}$ for devices 1, 2, 3 and 4
and $n_p\approx$6$\times$10$^{16}$~cm$^{-3}$ for device 5 were
estimated at room temperature by Hall measurement. Au/Cr
electrodes as source and drain were prepared for Ohmic contact by
sputtering. The channel intervals between the source and drain
were $L_{ch}$=3$\mu$m for devices 1, 2, and 3, $L_{ch}$=5$\mu$m
for device 5, and $L_{ch}$=10$\mu$m for devices 4, 6 and 7. The
thickness of the GaAs film for devices 1, 2, 3 and 4 is
approximately 350nm. Device 6 used an undoped GaAs film which
shows a $p$-type feature. The thickness of the GaAs film for
devices 5, 6, and 7 is approximately 100nm.

\begin{figure}
\vspace{-0.8cm}
\centerline{\epsfysize=9cm\epsfxsize=8cm\epsfbox{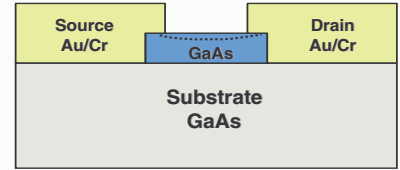}}
\vspace{-3.8cm} \caption{Schematic diagram of devices with two
terminals. The dotted line in the GaAs film is a channel (or path)
where current flows.}
\end{figure}

Electric characteristics of the devices were measured by a
precision semiconductor parameter analyzer (HP4156B). In the
measurements, a short pulse mode in HP4256B was used to prevent
Joule heat from occurring for a long pulse; the pulse width is
about 64${\mu}$sec. To protect devices from excess current, the
maximum current was limited to the compliance (or restricted)
current. The temperature dependence of Figs. 6 and 8 was measured
in a cryostat.

Micro-photoluminescence measurements were carried out with device
No. 6 at room temperature using an Acton spectrometer with a
liquid-nitrogen-cooled CCD camera and a high-NA ($\times$ 100)
corrected lens. The spectrally dispersed luminescence from
2$\mu$m-diameter regions between the source and drain electrodes
was observed. An Ar-ion laser of wavelength
 $\lambda\approx$488nm was used as the exciting light source.

\section{Results and Discussion}
Figure 3 shows the temperature dependence of resistance measured
for a bare GaAs film with a doping concentration of
$n_p\approx$5$\times$10$^{14}$~cm$^{-3}$. Resistance is an order
of 2$\times$10$^5~{\Omega}$ below 380 K and decreases
exponentially with increasing temperature above 380 K ; this
constitutes semiconducting behavior. This indicates that GaAs is a
valuable material for devices operating near room temperature. The
activation energy above 380 K is approximately 0.67eV. An abrupt
resistance change is not shown below 460 K because the temperature
of the structural phase transition of GaAs is approximately 1503 K
\cite{Gattow}. This is different from the temperature dependence
of resistance in VO$_2$ \cite{Kim1}.

\begin{figure}
\vspace{-0.8cm}
\centerline{\epsfysize=10cm\epsfxsize=8cm\epsfbox{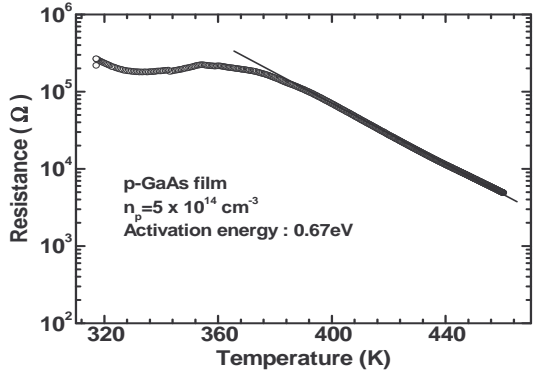}}
\vspace{-3.5cm} \caption{Temperature dependence of resistance for
a p-GaAs film with a hole doping concentration of
$n_p\approx$5$\times$10$^{14}$~cm$^{-3}$.}
\end{figure}

Figure 4 shows drain-source (DS) current, $I_{DS}$, and
drain-source voltage, $V_{DS}$, curves measured at room
temperature for device No. 1 with channel length
$L_{ch}$=3${\mu}$m and channel width $L_w$=50${\mu}$m. Abrupt
discontinuous current jump 1 and jump 2 are observed at
$V_{MIT}{\approx}$24V and $V_{MIT}{\approx}$-20V, respectively.
The derivative of a jump, $dJ/dV$, as shown in the inset,
corresponds to the density of states (DOS). Ohmic behaviors as
metal characteristics are exhibited above $V_{MIT}$s and are
caused by an internal resistance. Current density, J, just below
the Ohmic behavior at current 3.7 mA is approximately
7.5${\times}$10$^4$~A/cm$^2$. This value is much higher than that
(below 10$^2$ A/cm$^2$) measured in a semiconductor and
corresponds to an order of current density in a very dirty metal
due to internal resistance in the film and the effect of
measurement, which is described in a next section. The jump
corresponds to the jump of the effective mass of a quasiparticle
(or the increase of conductivity);
${\triangle}J/E{\equiv}{\triangle\sigma}{\propto}(m^{\ast}/m)^2$
in Eq. 1 where $E$ is an electric field at the jump and is the
same phenomenon as the jump measured in VO$_2$ \cite{Kim1}. Thus,
the jump of the DOS, the Ohmic characteristic, and the high
current density are characteristics of the first-order
metal-insulator transition. Asymmetry of jump 1 and jump 2 is
discussed in a next section.

Figure 5 shows hysteresis loops observed by $I_{DS}$-$V_{DS}$
curve measurements for device No. 2 with $L_{ch}$=3$\mu$m.
Measurement was performed in a series of "start $\rightarrow$ 1
$\rightarrow$ 2 $\rightarrow$ 3 $\rightarrow$ 4 $\rightarrow$
end". Double hysteresis at jumps 1 and 4 and jumps 2 and 3 are
exhibited. Jump 1 ( or jump 3) and jump 2 (or jump 4),
respectively, are asymmetrical in magnitude of jumps and V$_{MIT}$
of jumps, but measurement ($\circ$) of jumps 1 and 2 and
measurement ($\bullet$) of jumps 3 and 4 are symmetrical. Jump 2,
the transition from insulator to metal, results from doping a low
concentration of hole $n_c$ \cite{Yslim} and is not accompanied by
heat, whereas jump 3, the transition from metal to insulator, is
accompanied by Joule heat produced by scattering of carriers in
metal. The Joule heat causes an extra excitation, which results
from doping a very small hole concentration $\alpha$;
$n_c>>\alpha$. That is, jump 2 is due to doping of $n_c$ and jump
3 is attributed to doping of $n_c$ + $\alpha$. On the basis of the
hole-driven MIT theory, the magnitude of the jump and V$_{MIT}$
when $n_c$ + $\alpha$ is doped are less than those when $n_c$ is
doped; this causes asymmetry. Thus, the observation of hysteresis
indicates that the abrupt jump for GaAs is evidence of the
first-order MIT.

\begin{figure}
\vspace{-0.8cm}
\centerline{\epsfysize=9cm\epsfxsize=8cm\epsfbox{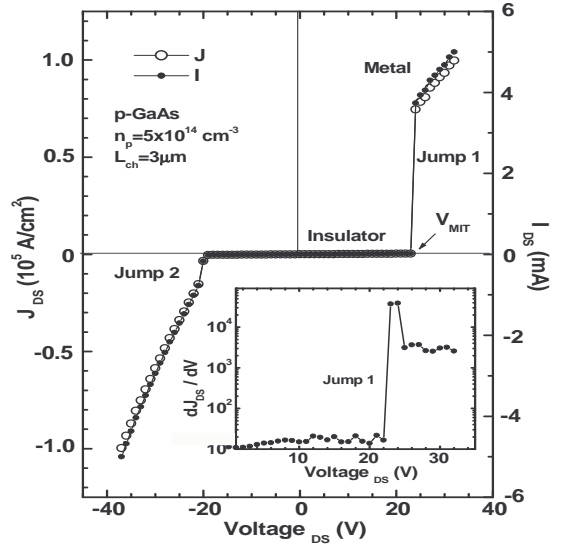}}
\vspace{0.0cm} \caption{I$_{DS}$-V$_{DS}$ curve measured for
device No. 1 fabricated with a p-GaAs film with a hole doping
concentration of $n_p\approx$5$\times$10$^{14}$~cm$^{-3}$. The
first and the second measurements are denoted by circle ($\circ$)
and solid ($\bullet$), respectively. The inset shows a derivative,
dJ/dV, regarded as the density of states.}
\end{figure}

Figures 6a and 6b show the temperature dependence of
$I_{DS}$-$V_{DS}$ curves measured with device No. 3. With
increasing temperature, leakage current at points A, B, C, and D
increases, MIT-$V_{DS}$ decreases, MIT-$I_{DS}$ increases, and the
magnitude of the jump decreases, as indicated by MIT points E, F,
G, and H, as shown in Fig. 6a. This is attributed to an increase
of the excitation of holes with increasing temperature. As
outlined in the previous section in relation to Fig. 4, the
temperature dependence is explained by the hole-driven MIT theory.
Furthermore, a line just below the jump measured at 350 K
indicates an exponential increase with increasing an applied
voltage. This is a semiconducting behavior and suggests that the
abrupt MIT occurs by way of the semiconduction as mentioned in a
previous paper \cite{Kim1}.

\begin{figure}
\vspace{-0.8cm}
\centerline{\epsfysize=9cm\epsfxsize=8cm\epsfbox{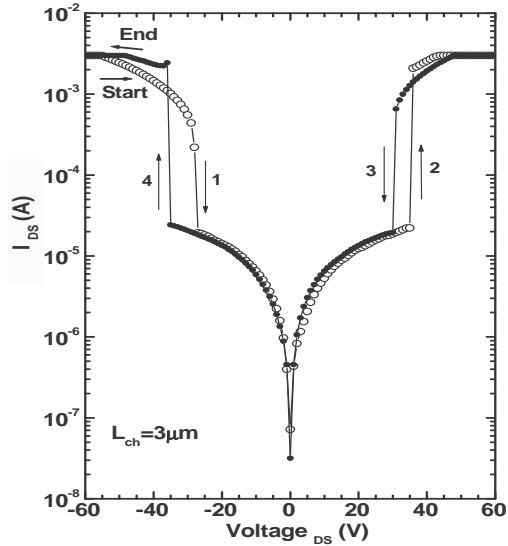}}
\vspace{-0.3cm} \caption{Hysteresis loops measured for device No.
2 fabricated with a p-GaAs film with a hole doping concentration
of $n_p\approx$5$\times$10$^{14}$~cm$^{-3}$.}
\end{figure}
\begin{figure}
\vspace{-0.8cm}
\centerline{\epsfysize=9cm\epsfxsize=8cm\epsfbox{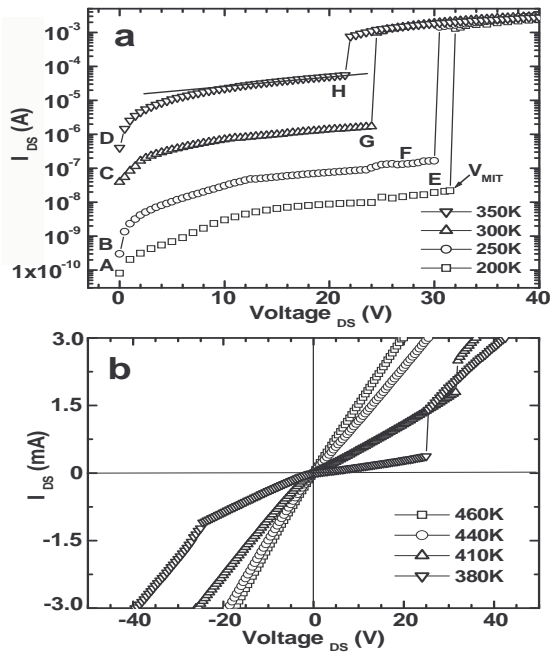}}
\vspace{0.5cm} \caption{Temperature dependence of
I$_{DS}$-V$_{DS}$ curve measured for device No. 3 fabricated with
a p-GaAs film with a hole doping concentration of
$n_p\approx$5$\times$10$^{14}$~cm$^{-3}$.}
\end{figure}

The temperature dependence also provides decisive information for
revealing the mechanism of the abrupt jump. When the number of
total holes, $n_{tot}$, in the hole levels is given by
$n_{tot}=n_b + n_{free}(T,E)$, where $n_b$ is the number of bound
holes in the levels and $n_{free}(T,E)$ is the number of holes
freed by temperature $T$ and electric field $E$ from the levels,
$n_b$ decreases with increasing $n_{free}(T,E)$. For the abrupt
jump, ${\triangle}n{\equiv}n_c-n_{free}$=0 should be satisfied. At
$T_{tr}\approx$440 K determined in the next section, it is
suggested, as decisive evidence of the Mott transition, that the
abrupt current jump disappears, because $n_{free}(T{\approx}440K,
E{\approx}0)$=$n_c$ ($i.e.~{\triangle}n$=0) is excited only by
temperature. Below $T_{tr}\approx$440 K, the abrupt MIT voltage
decreases with increasing temperature, because, from
$n_c{\equiv}n_{free}(T,E)$=$n_{free}(T)$ + $n_{free}(E)$, the
increase of $n_{free}(T)$ with increasing temperature decreases
$n_{free}(E)$. Thus, it is revealed that the mechanism of the
abrupt MIT excited by temperature is the same as that by an
electric field. Temperature and electric field are a means of
exciting hole charges.

\begin{figure}
\vspace{-0.8cm}
\centerline{\epsfysize=9cm\epsfxsize=8cm\epsfbox{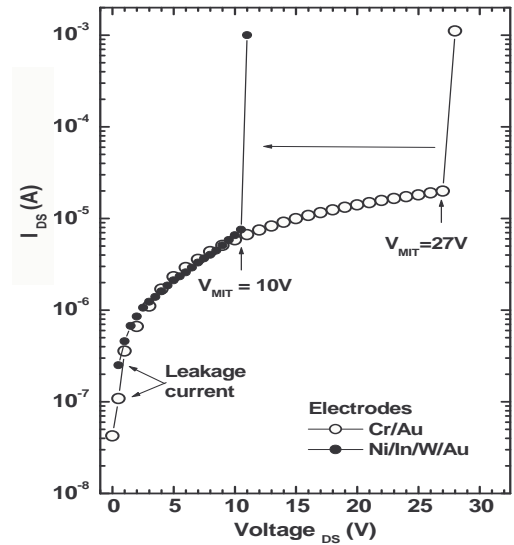}}
\vspace{-0.3cm} \caption{Electrode dependence of I$_{DS}$-V$_{DS}$
curve measured for two-terminal devices fabricated with the same
configuration of $L_{ch}$=5$\mu$m, $L_{w}$=50$\mu$m over a p-GaAs
film with a hole doping concentration of
$n_p\approx$5$\times$10$^{14}$~cm$^{-3}$. Both Au/W/In/Ni (solid
bold) and Au/Cr (circle) were used for source and drain
electrodes. Compliance current of 1 mA was applied to protect the
device.}
\end{figure}

On the other hand, we also investigate the trend of
$I_{DS}-V_{DS}$ curves at higher temperature. Fig. 6b shows the
increase of MIT-$V_{DS}$ at 380 K and 410 K, the decrease of the
magnitude of jump, and a large increase of slope of the Ohmic
behavior below jump; this is quite a different new phenomenon from
that seen in Fig. 6a. This appears to be attributed to thermal
excitation from bound electrons in sub-band of energy gap 1.45eV
into carriers, because the excitation probability of bound
electrons across the energy gap increases with increasing
temperature. The jump is not observed above 440 K. The Ohmic
behaviors at 440 K and 460 K are similar. Thus, GaAs undergoes a
MIT near 440 K and shows metallic characteristics above 440 K
which is far below the structural-phase-transition temperature of
1503 K. Thus, the MIT is far from the structural change.

\begin{figure}
\vspace{-0.8cm}
\centerline{\epsfysize=9cm\epsfxsize=8cm\epsfbox{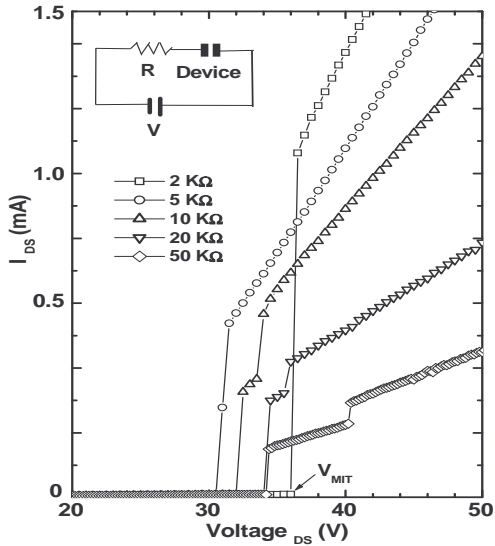}}
\vspace{-0.3cm} \caption{Resistance dependence of resistance
measured for device No. 4 fabricated with a p-GaAs film with a
hole doping concentration of
$n_p\approx$5$\times$10$^{14}$~cm$^{-3}$.}
\end{figure}

Figure 7 shows a large decrease of MIT-$V_{DS}$ from $V_{MIT}$=27V
to 10V and additional increase of leakage current near zero
voltage for a device patterned with electrodes of source and drain
of Au/W/In/Ni (instead of Au/Cr), which had lower resistance on
contact surface with a GaAs film. This phenomenon is similar to
the temperature dependence of the MIT in Fig. 6a and denies that
the abrupt MIT (or jump) occurs only by very high electric field.

Figure 8 shows the resistance dependence of $I_{DS}$-$V_{DS}$
curves measured with device No. 4. Resistance is connected in a
serial configuration with the device, as shown in the inset of
Fig. 8. With increasing external resistance, the magnitude of the
current jump decreases and $V_{MIT}$ decreases by 5K$\Omega$ and
again increases from 5K$\Omega$ to 50K$\Omega$. The increase of
external resistance is interpreted as an increase of the insulator
region (region A in Fig. 1b) not undergoing the abrupt MIT in the
measurement region. As for the increase of $V_{MIT}$ with
increasing resistance from 5K$\Omega$ to 50K$\Omega$, since the
increase of region A in the right side of Fig. 1b decreases
conductivity $\sigma$, $V_{MIT}$ increases when $J$'s are constant
at V$_{MIT}$s and the magnitude of the jump decreases when $E$ is
constant; $\sigma$=$J/E$ and $E=V/d$ where $d$ is channel length.
Note that $J$s at V$_{MIT}$s from 5K$\Omega$ to 50K$\Omega$ had
nearly similar values in the logarithmic picture although
V$_{MIT}$s are not shown in Fig. 8. Furthermore, step jumps in the
$I_{DS}-V_{DS}$ curves of 10K$\Omega$, 20K$\Omega$ and 50K$\Omega$
may be due to inhomogeneity in the GaAs film.

\begin{figure}
\vspace{-0.8cm}
\centerline{\epsfysize=9cm\epsfxsize=8cm\epsfbox{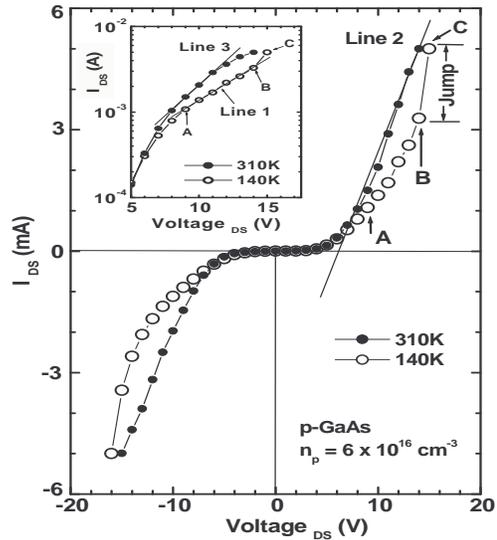}}
\vspace{-0.5cm} \caption{I$_{DS}$-V$_{DS}$ curves measured for
device No. 5 fabricated with a p-GaAs film with a hole doping
concentration of $n_p\approx$6$\times$10$^{16}$~cm$^{-3}$.}
\end{figure}

We take into account that internal resistance in a GaAs film makes
it possible to observe Ohmic behavior, as shown in Figs. 4, 5, and
6. The Ohmic behavior indicates that channel layers in devices (or
GaAs) are inhomogeneous. The magnitude of an observed jump is not
an intrinsic value but the effect of measurement and increases
with decreasing the internal resistance. When only the metal
region in Fig. 1b is measured, the magnitude of the current jump
might be of an order of 10$^{7{\sim}8}$ A/cm$^2$, the current
density of a good metal. This indicates that the observed $I_{DS}$
changes with external resistance, even though the intrinsic metal
characteristics remain unchanged. Thus, the observed current
density, $J_{DS}$, of an order of 10$^{4{\sim}5}$ A/cm$^2$, as
shown in Fig. 4, is merely an average of the metal region over the
measurement region. The average is the effect of measurement.
Thus, a true current jump cannot be measured in an inhomogeneous
system, as suggested by the hole-driven MIT theory
\cite{Kim1,Kim2}.

Figure 9 shows the temperature dependence of $I_{DS}$-$V_{DS}$
curves measured with device No. 5 fabricated by a p-GaAs film with
$n_p{\approx}6{\times}$10$^{16}~$cm$^{-3}$ larger than $n_c$. In
the curve measured at 140 K, data between point A and point B fit
a linear in the logarithmic plot, as shown by line 1 in the inset,
which indicates the semiconducting behavior. A small jump between
point B and point C corresponds to the jump of the abrupt MIT. The
small jump is due to the increase of an internal resistance by a
larger doping than $n_c$, as shown in Fig. 8. In the curve
measured at 310 K, line 2 without jump denotes the Ohmic metallic
behavior rather than an exponential semiconducting behavior, as
shown by line 3 in the inset; line 3 does not fit all data above
V$_{DS}$=8V in the logarithmic plot. This indicates that, when a
much higher hole concentration than $n_c$ is doped into the
valence band, a continuous MIT instead of an abrupt MIT occurs due
to the decrease of the magnitude of the jump, as predicted in the
hole-driven MIT theory. That is, disappearance of the small jump
at 310 K are because $n_{free}(E)$ related to the jump at 140 K
became zero at 310 K and $n_{free}(T)$ at 310 K was a larger than
$n_c$ (see Fig. 6). Therefore, the continuous MIT is not intrinsic
and may be MITs observed in MOSFET and GaAs/AlGaAs structures
\cite{Gold,Yoon,Pudalov,Shashkin}. Furthermore, it is explained
why an semiconducting effect is not observed above V$_{DS}$=8V in
the I-V curve measured at 310 K. Since region A in Fig. 1b plays a
role of resistance to region B (metal region) after the abrupt MIT
and acts as a semiconductor with an exponential current
characteristic in the I-V curve, the magnitude of current measured
in relatively smaller region A (semiconductor) than region B is
much smaller than that in region B (metal) in the high electric
field above the MIT-V$_{DS}$; a measured conductivity
$\sigma_{measured}=\sigma_{semiconductor}+\sigma_{metal}\approx\sigma_{metal}$
because $\sigma_{semiconductor}<<\sigma_{metal}$ after the MIT
V$_{DS}$.

\begin{figure}
\vspace{-0.8cm}
\centerline{\epsfysize=9cm\epsfxsize=8cm\epsfbox{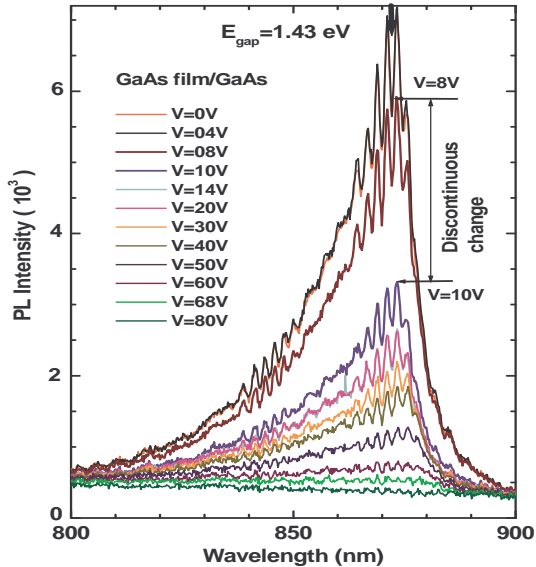}}
\vspace{-0.0cm} \caption{Photoluminescence (PL) spectra measured
for device No. 6 fabricated with an un-doped GaAs film/GaAs
substrate.}
\end{figure}

Figure 10 shows photoluminescence (PL) spectra measured by an
Ar-ion laser light of wavelength 488nm with device No. 6
fabricated by using an undoped GaAs film on GaAs substrate, in
order to observe the change in the energy gap of GaAs. Device No.
6 displays a clean abrupt jump characteristic in I-V curve
($\circ$) in Fig. 11a. The energy gap is approximately 1.43 eV.
The intensity of spectra decreases as V$_{DS}$ increases, and
disappears at V$_{DS}$=80V. In particular, a discontinuous change
in the intensity of spectra between V$_{DS}$=8V and V$_{DS}$=10V
($\bigtriangleup$V$_{DS}$=2V) is observed.
$\bigtriangleup$V$_{DS}$ corresponds to a change of voltage while
a MIT jump in an I$_{photocurrent}$-V curve ($\bullet$) occurred,
as shown in Fig. 11a; the PL spectra and the I$_{photocurrent}$-V
curve ($\bullet$) were measured simultaneously in an exposed state
of laser light. Note that, as shown in Fig. 11, when device No. 6
was exposed to the laser light, a considerable amount of
photocurrent due to photoexcited holes and electrons occurred; as
a result, V$_{MIT}$ decreased due to $n_c$ as indicated by arrow
A, the current (or photocurrent) increased as indicated by arrow
B, and the MIT jump decreased largely as indicated by arrow C.

\begin{figure}
\vspace{-0.8cm}
\centerline{\epsfysize=9cm\epsfxsize=8cm\epsfbox{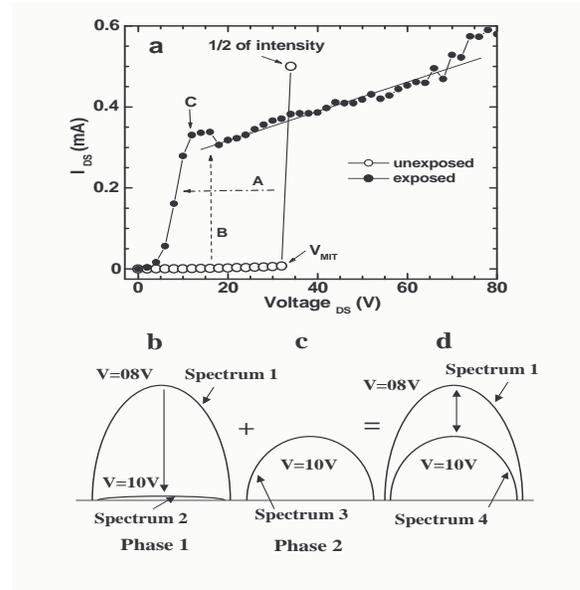}}
\vspace{-0.5cm} \caption{{\bf a} I$_{DS}$-V$_{DS}$ curves measured
for device No. 6 fabricated with an un-doped GaAs film/GaAs
substrate. Exposed I-V curve ($\bullet$) measured in Ar-ion laser
light with wavelength 488nm shows a photocurrent effect. Unexposed
I-V curve ($\circ$) is given as a reference and half of the true
I$_{DS}$ value in the figure. The current was restricted by a
compliance current of 1mA so as to protect the device. {\bf b, c,
d} Schematic figures explaining the discontinuous change of the
spectra in Fig. 10. Spectrum 4 is an average of spectrum 2 and
spectrum 3.}
\end{figure}

The discontinuous change in Fig. 10 can be interpreted on the
basis of the hole-driven MIT theory as follows. Since the abrupt
MIT jump in a homogeneous Mott insulator is caused by hole doping
of a low concentration, the undoped GaAs becomes inhomogeneous and
has two phases after the abrupt MIT, as shown in Fig. 1b. One
phase (phase 1) undergoing the abrupt MIT is a homogeneous Mott
insulator, but the other phase (phase 2) not undergoing the abrupt
MIT is an insulator (or semiconductor at room temperature). First,
below V$_{DS}$=8V, because phase 1 is a Mott insulator, the PL is
observed as one spectrum, i.e., as spectrum 1 in Fig. 11b. Second,
at V$_{DS}$=10V, because phase 1 undergoes the abrupt MIT,
spectrum 1 becomes spectrum 2 without the energy gap, as shown in
Fig. 11b, and because phase 2 appears after the MIT, a new
spectrum (spectrum 4) as spectrum 3 appears as shown in Fig. 11c.
The new spectrum is an average of spectrum 2 and spectrum 3 and is
nearly the same as spectrum 3, because spectrum 2 is nearly too
small. The intensity of the spectrum 3 can be smaller than
spectrum 1, because region A in Fig. 1b is smaller than region B
and decreases little by little with increasing $V_{DS}$. Spectrum
3 is regarded as a pseudogap. Third, phase 2 finally undergoes a
continuous MIT in a high applied voltage (or electric field), as
shown in Fig. 10, and is regarded as an internal resistance (or
insulating) phase causing Ohmic behaviors in Figs 4 and 6. At the
applied voltages of $V_{DS}$=8V and $V_{DS}$=10V, the
corresponding spectra can be represented as shown in Fig. 11d.
Thus, the discontinuous change of spectra is attributed to the
extinction of the energy gap in phase 1 due to the abrupt MIT.
Conversely, when the abrupt MIT does not occur, the spectra in
Fig. 9 monotonously decrease without a discontinuous change.
Furthermore, the pseudogap is similar to pseudogaps observed in
high-$T_c$ superconductors \cite{Blanton,Kim4}.

\begin{figure}
\vspace{-0.3cm}
\centerline{\epsfysize=9cm\epsfxsize=8cm\epsfbox{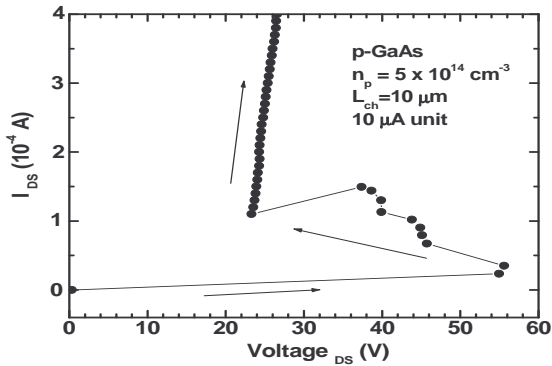}}
\vspace{-3.3cm} \caption{Negative differential resistance measured
for device No. 7 fabricated with p-GaAs film with a hole doping
concentration of $n_p\approx$5$\times$10$^{14}$~cm$^{-3}$.}
\end{figure}

As further MIT evidence, the negative differential resistance
(NDR) was also measured by the current-controlled method by device
No. 7 for p-GaAs film with
$n_p{\approx}5{\times}$10$^{14}~$cm$^{-3}$. However, the NDR was
not observed in devices fabricated with p-GaAs film with
$n_p{\approx}6{\times}$10$^{16}~$cm$^{-3}$. The NDR is regarded as
an MIT due to doping of $n_c$ holes excited by an increase of
temperature caused by current, as in the case of VO$_2$
\cite{Stefanovich}. The mechanism of the NDR is the same as that
of electric-field induced MIT due to $n_c$, although the
excitation method is different.

The abrupt first-order MIT observed in GaAs, not undergoing a
structural phase transition, is a new phenomenon that cannot be
explained by well-known semiconductor theories or MIT theories
except the Brinkman-Rice (BR) picture and the extended BR picture
(Hole-driven MIT theory) with singularity. The MIT is the same as
that observed in VO$_2$. Thus, the abrupt MIT observed in the
devices with p-GaAs film is different from continuous MITs, in the
2D systems \cite{Gold,Yoon,Pudalov,Shashkin,Zhu,Lilly,Leturcq},
which may be due to over-doping rather than $n_c$ or it may be a
metallic behavior by an excitation of bound charges in impurity
levels below the abrupt jump in Fig. 6.

Furthermore, on the basis of the hole-driven MIT theory, the
abrupt first-order MIT can occur even in other semiconductors with
both an energy gap of less than 2 eV and hole levels and parent
materials BaBiO$_3$, La$_2$CuO$_4$, and PrBaCuO$_7$ of high-$T_c$
superconductors and parent material LaMnO$_3$ of a colossal
magnetoresistance. In a semiconductor such as InP/InGaAs/InP
substrate with doped electrons of very low concentration, an
abrupt MIT (or jump) was not observed.

\section{Comparison of abrupt MIT and breakdown}
The jump in the abrupt MIT is similar to a phenomenon "avalanche
breakdown" which is caused by a high electric field in a $pn$
junction device such as a Zener diode, but doesn't show the
metallic behavior \cite{Brophy,Sze}. The breakdown is a longtime
unsolved problem in semiconductor physics; what is the identity of
the breakdown and why can the breakdown happen to be a jump even
though avalanche phenomenon is continuous \cite{Oliver,Kolodzey}.
The breakdown causes a degradation of film quality or device
breakdown, which might occur due to Joule heating arising from a
trapped high current ($J{\approx}\sim$10$^{7{\sim}8}$~A/cm$^2$) at
defects just after the abrupt MIT happened. Its main
characteristic is to remove repeatability of data in experiments.
That is, the breakdown does not take place in as clean films as
devices used in this research; for example, pure metals does not
suffer from breakdown even if high current flows. Likewise, the
breakdown ahead of the abrupt MIT is hard to occur for a
stoichiometric GaAs with less impurities; we experimentally
checked the absence of the abrupt MIT (or breakdown)in GaAs in the
even harsh conditions, 20MV/m and 100$^{\circ}$C and, as a result
of the experiment, the semiconducting behavior was observed.
Moreover, we have confirmed that there was no even trace of an
abrupt MIT for n-type semiconductor such as InP substrate. This
can be explained not in view of breakdown but by the fact that the
abrupt jump does not occur due to the absence of the critical hole
doping $n_c$, as the MIT theory suggests. Because the breakdown is
not theoretically well-established, it does not account
consistently for the Ohmic behavior above the jump, hysteresis,
temperature dependence of the jump, electrode dependence of the
jump, resistance dependence of the jump, doping dependence of the
jump, negative differential resistance, breakdown of the energy
gap, as shown in this paper. On the contrary, the abrupt MIT
theory can explain the breakdown phenomenon.

\section{Conclusion}
An abrupt first-order MIT for GaAs occurs by hole doping of very
low concentration, irrespective of the critical magnitude of the
applied electric field, and is accompanied with inhomogeneity. The
MIT does not undergo a structural phase transition. Continuous MIT
takes place at a larger hole concentration than a critical hole
doping $n_c$ in GaAs. A self-consistent theory needs to be
developed to explain the abrupt MIT in GaAs irrespective of a band
insulator. Furthermore, the abrupt MIT will be very valuable for
new multi-functional devices for next generation.

\section*{ACKNOWLEDGEMENTS}
This research was performed by a project of "High-Risk
High-Return" supported by ETRI. YS Lim measured photoluminescence
spectra and was supported by Konkuk University to provide optical
measurement tools for this research.






\begin{references}
\bibitem{Yslim} For GaAs, effective mass $m^{\ast}/m$=0.082,
dielectric constant $\varepsilon$=12, Bohr radius a$_B$ =
a$_B$({hydrogen})$\varepsilon$/($m^{\ast}/m$) = 0.53*12/0.082 =
77$\AA$. From Mott criterion $n_c^{1/3}a_B\approx$0.25,
$n_c\approx$3.3${\times}$10$^{16}~$cm$^{-3}$, when it is assumed
that one electron exists in the unit cell.
\end{references}
\end{document}